\documentclass[aps,prb,reprint,twocolumn,superscriptaddress]{revtex4-2}
\usepackage{amssymb,amsmath,amsfonts}
\usepackage{graphicx} 
\usepackage{float} 
\usepackage{natbib} 
\usepackage{xcolor}
\usepackage[colorlinks=true,bookmarks=false,citecolor=blue,urlcolor=blue]{hyperref}
\usepackage{lmodern}
\usepackage{soul}

\graphicspath{ {Figures/} }

\newcommand{\eps}{\varepsilon}
\newcommand{\w}{\omega}

\newcommand{\RH}{ \mathrm{RH} }
\newcommand{\LH}{ \mathrm{LH} }

\newcommand{\sinc}{\mathbf{s}^+ }
\newcommand{\sout}{\mathbf{s}^- }

\newcommand{\moscow}{
Moscow Center for Advanced Studies, Moscow 123592, Russia
}

\newcommand{\skolkovo}{
Center for Engineering Physics, Skolkovo Institute of Science and Technology, Moscow, 121125, Russia.
}

\newcommand{\xpanceo}{
Emerging Technologies Research Center, XPANCEO, Internet City, Emmay Tower, Dubai, United Arab Emirates
}

\begin{document}

\title{Low-symmetry lattices of non-chiral meta-atoms for resonant handedness-preserving reflection}

\author{Anastasia Pozharkova}
\affiliation{\moscow}
\affiliation{\skolkovo}

\author{Oleg Blokhin}
\affiliation{\moscow}
\affiliation{\skolkovo}

\author{Sergey A. Dyakov}
\affiliation{\skolkovo}

\author{Denis G. Baranov}
\email[]{baranov.mipt@gmail.com}
\affiliation{\moscow}
\affiliation{\skolkovo}
\affiliation{\xpanceo}

\begin{abstract}
Mirrors that preserve the handedness of optical radiation upon reflection are an essential building block for the design of numerous resonant nanophotonic structures with capabilities for enantiomeric discrimination. 
Ordinary metallic and Bragg dielectric mirrors are not suitable in these context since they flip handedness of electromagnetic field upon reflection around normal incidence.
While there has been considerable progress in the development of such reflecting structures, this research area remains largely unexplored.
Here, we present a detailed numerical and theoretical analysis of the potential of low-symmetry periodic lattices composed of high-symmetry non-chiral meta-atoms for resonant reflection with handedness preservation (HP).
Using full-wave numerical simulations, we analyze a family of rhombic and monoclinic (oblique) lattices of circular dielectric disks and/or holes, and in each identify the regime of near-perfect HP reflection.
We study the robustness of these structures to geometric deviations, material losses, and incidence angle. 
Finally, we describe the resonant HP response of these structures using the coupled-mode theory formalism.
\end{abstract}

\maketitle
\newpage

\section{Introduction}


Various optical mirrors are an essential building blocks for constructing Fabry-Perot optical cavities \cite{pfeifer2022achievements, sorger2009plasmonic, hunger2010fiber}, which are further used for enhancing light-matter interaction  with electronic transitions of quantum emitters \cite{Lodahl2015}, or vibrational transitions of solids \cite{canales2024self}. 
Approaches to optical mirrors range from ordinary metallic films \cite{Munkhbat2021} to dielectric Bragg reflectors \cite{wang2005plasmon, kawasaki1978narrow} and more complicated reflecting metasurface with tailored spectral response \cite{ding2017gradient, park2017dynamic, arbabi2017planar} and even reflecting arrays made of a collection of atoms \cite{Rui2020, Merchiers2007}.

Most of these approaches, while offering great reflecting performance, preserve the linear polarization states of light upon reflection at normal incidence. 
And while in most scenarios this is exactly what is needed, there are situations when one wants to reflect electromagnetic radiation with  preservation of \emph{handedness} instead, which requires a complete conversion of circular polarization (spin) of the plane wave \cite{Fernandez-Corbaton2016, baranov2020circular}.
This is, for example, the case for so called handedness-preserving (HP) cavities, which are intended to enhance the light-matter interaction of optical radiation with resonances of chiral media \cite{Condon1937, Valev2013}, whose electronic transitions combine parallel electric and magnetic dipole moments \cite{Tang2010, Govorov2010, Klimov2012, Yang2009, Yang2011}.

Architectures based on metallic \cite{Plum2015} and dielectric \cite{Semnani2020, duan2024spin} metasurfaces with complicated unit cells have been mostly studied and used for this purpose.
One should also mention numerous metasurface-enabled approaches for polarization control in \emph{transmission} \cite{Gansel2009, Zhao2011, Zhao2012, ding2015ultrathin, Kang2015, BalthasarMueller2017, Cerjan2017, Dorrah2021}.

The progress in the design and analysis of such metasurfaces has lead to the development of various HP optical cavities \cite{Feis2020, Voronin2022, dyakov2023chiral, Dams2025} and theoretical predictions of various intriguing effects involving chiral molecular enantiomers \cite{Beutel2021, mauro2023chiral, baranov2023toward, schafer2023chiral, Riso2023, Ke2023, Graf2019, Dyakov2025}.
However, these structures often require fabrication of complicated geometries that may introduce a lot of uncertainty. 
This calls for the search for simpler approaches towards design and fabrication of chiral and HP structures.
One such example is presented by van-der Waals enabled lithography-free fabrication of multilayers with significant chiro-optical response \cite{voronin2024chiral}.
Unlike previous chiral metasurfaces that rely on structurally chiral unit cells, Toftul et al. recently demonstrated that it is possible to engineer optical chirality in \emph{monoclinic} lattices of non-chiral dielectric meta-atoms by exploiting the lattice symmetry alone  \cite{Toftul2024, Toftul2025, Sinev2025}.

Here, we follow a similar strategy and demonstrate the emergence of HP reflection (HPR) at normal incidence in rhombic and monoclinic lattices of simple circular disks and holes.
We analyze the performance of a mirror-symmetric rhombic lattice for symmetric right-handed-to-right-handed (RH) and left-handed-to-left-handed (LH) near-perfect HPR, and inspect different architectures based on an array of circular dielectric disks and array of circular holes in a dielectric membrane.
We study the effect of material absorption on HPR, and explain the origin of the HPR response in the rhombic structures using the coupled-mode theory.
Next, we violate vertical plane of mirror symmetry by considering a monoclinic lattice of scatterers, which allows near-perfect reflection of only one handedness.
In a similar fashion, we analyze the robustness of the monoclinic structure, the effect of material absorption, and the validity of coupled-mode theory.
Our results provide a new class of structures and design rules for HP metasurface mirrors.

\begin{figure*}[t!]
\centering\includegraphics[width=1.0\textwidth]{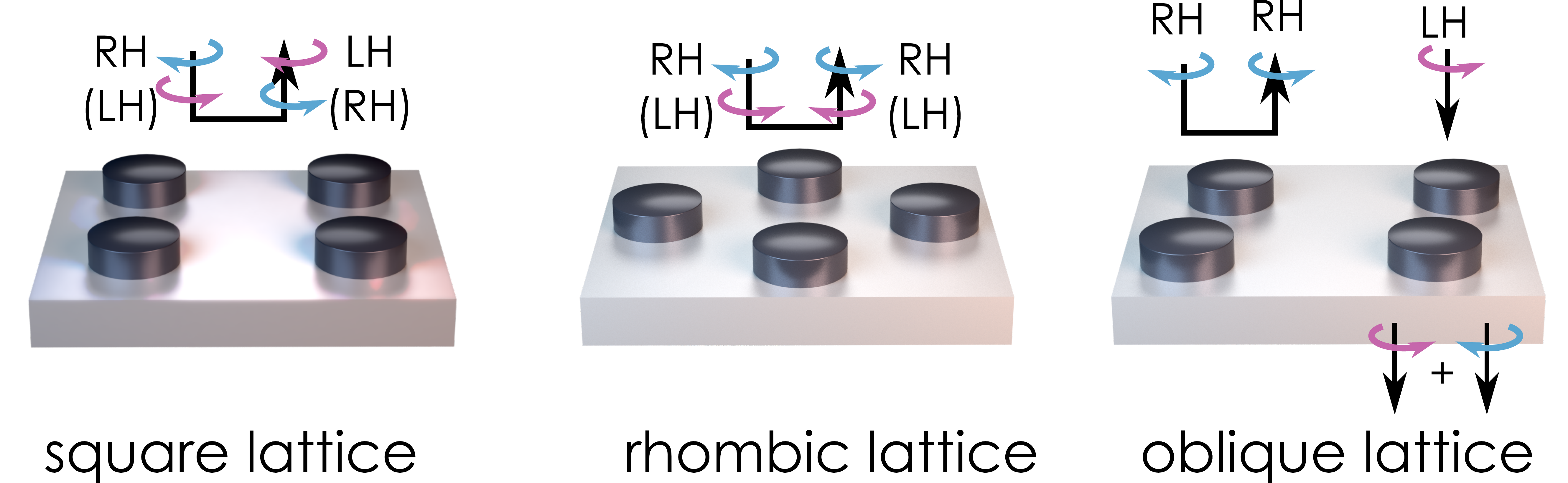}
\caption{Allowed polarization conversion processes in reflection from a sub-diffraction metasurface in circular polarization basis.
(a) Sub-diffraction $C_{4v}$ symmetric square array of circular disks exhibits only cross-polarized reflection at normal incidence.
(b) Sub-diffraction $C_{2v}$ symmetric rhombic array of circular disks in addition exhibits equal/symmetric co-polarized reflection at normal incidence.
(c) Sub-diffraction $C_{2}$ symmetric oblique (monoclinic) array of circular disks (is capable of) exhibits unequal/asymmetric co-polarized reflection at normal incidence.
}
\label{fig1}
\end{figure*}

\section{Results}

Figure \ref{fig1} shows the geometries of periodic lattices analyzed in our study: we analyze a periodic lattice of dielectric circular disks in air or on a substrate.
Most studies of metasurfaces focus on square or rectangular arrays of meta-atoms \cite{landy2008perfect, meinzer2014plasmonic, Gorkunov2020, zograf2025ultrathin}.
A square lattice of simple circular disks possesses $C_{4v}$ symmetry and offers no spin conversion at normal incidence, Fig. \ref{fig1}(a) \cite{Menzel2010, Kruk2015}.
To enable handedness preservation upon reflection accompanied by spin flipping, we lower the symmetry of the lattice by considering a rhombic lattice with $C_{2v}$ or $D_{2h}$ symmetry, Fig. \ref{fig1}(b).
This allows reflection with handedness conservation. 
Nonetheless, $\RH \to \RH$ and $\LH \to \LH$ reflections will occur with equal efficiencies from such a metasurface due to the presence of a vertical plane of mirror symmetry \cite{dyakov2023chiral}.
To enable \emph{unequal} $\RH \to \RH$ and $\LH \to \LH$ reflections at normal incidence, we further lower the symmetry to $C_2$ or $C_{2h}$ by considering an oblique lattice of disks, which at most has an in-plane mirror symmetry (thus rendering the whole structure non-chiral) and allows for unequal $\RH \to \RH$ and $\LH \to \LH$ reflections, Fig. \ref{fig1}(c).

\subsection{Rhombic lattices}

We begin our analysis of HPR effect in various families of rhombic lattices. The symmetry of the structure ensures the absence of geometrical chirality and equal $\RH \to \RH$ and $\LH \to \LH$ reflections:
\begin{equation}
    r_{RR} = r_{LL}.
\end{equation}
Thus we optimize the performance of rhombic metasurfaces in terms of the HP intensity reflection coefficient $|r_{RR}|^2$.

Two-dimensional rhombic lattice is defined by the real-space translation vectors
\begin{equation}
\mathbf{a}_1 = L(1,0), \qquad
\mathbf{a}_2 = L(\cos\phi,\;\sin\phi),
\end{equation}
where $L$ is the lattice period and $\phi$ is the angle between the basis vectors. The corresponding reciprocal lattice vectors are given by
\begin{equation}
    \mathbf{b}_1 = \frac{2\pi}{L\sin\phi}\,(\sin\phi,\,-\cos\phi), \qquad
    \mathbf{b}_2 = \frac{2\pi}{L\sin\phi}\,(0,\;1),
\end{equation}
Magnitudes of the few relevant lowest-order reciprocal vectors are given by:
\begin{equation}
\begin{split}
    |\mathbf{G}_{\pm1,0}| &= |\mathbf{G}_{0,\pm1}| = \frac{2\pi}{L\sin\phi}, \\
    |\mathbf{G}_{1,1}| &= |\mathbf{G}_{-1,-1}| = \frac{2\pi}{L\sin\phi}\sqrt{2(1-\cos\phi)}= \frac{2\pi}{L\cos\frac{\phi}{2}}, \\
    |\mathbf{G}_{1,-1}| &= |\mathbf{G}_{-1,1}| = \frac{2\pi}{L\sin\phi}\sqrt{2(1+\cos\phi)}= \frac{2\pi}{L\sin\frac{\phi}{2}}.
\end{split}
\end{equation}

The following analysis will be restricted to the sub-diffraction spectral range, where the transmitted and reflected light is represented by a single propagating diffraction channel; i.e., the array operates in the metasurface regime \cite{Xu2016Planar, gomez2006extraordinary, DeAbajo2007}.
At normal incidence ($\mathbf{k}_{\parallel}=0$) the diffraction channel $(m,n)$ is open for radiation if:
\begin{equation}
    |\mathbf{k}_{\parallel} + \mathbf{G}_{m,n}| \le k_0,
\end{equation}
Thus, at normal incidence the sub-diffraction regime corresponds to frequencies below this threshold:
\begin{equation}
    \w/c < G_\mathrm{min},
\end{equation}
where
\begin{equation}
    G_\mathrm{min} \equiv \min_{(m,n)\neq(0,0)} |\mathbf{G}_{m,n}|
\end{equation}
denotes the smallest non-trivial reciprocal lattice vector of the rhombic lattice.

\subsubsection{Array of circular disks in air}

\begin{figure*}[t!]
\centering\includegraphics[width=.9\textwidth]{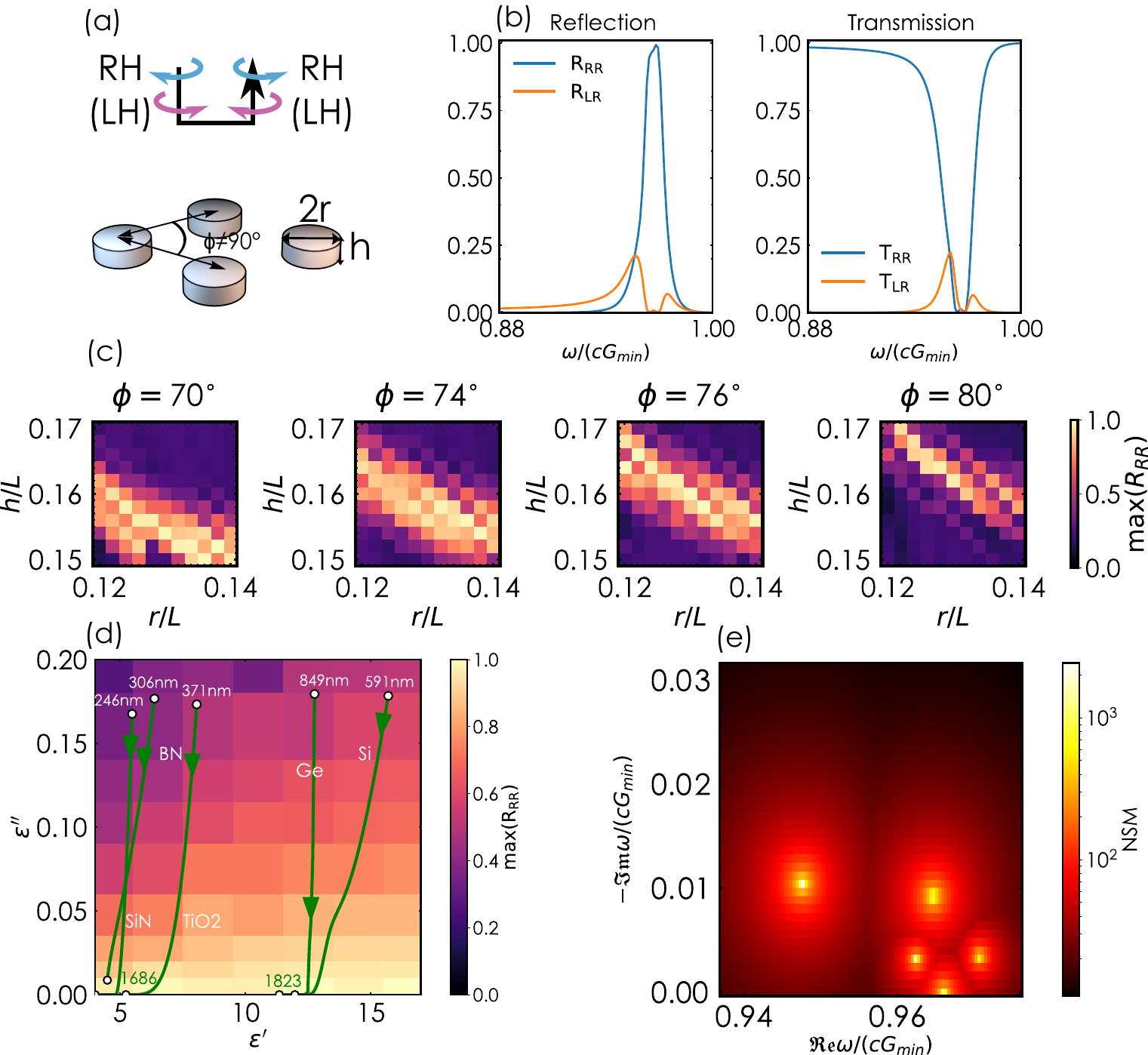}
\caption{HPR in a rhombic lattice of circular disks in air. 
(a) A sketch of the geometry of the system.
(b) Numerically obtained reflection and transmission [circularly-polarized basis] spectra of an optimal rhombic lattice of dielectric ($n = 4$) disks in the circular polarization basis. 
The data is presented as a function of normalized frequency $\w/(cG_{\min})$.
(c) Density plots of maximal achievable HPR ($|r_{RR}|^2$) within the sub-diffraction spectral range ($\w/cG_{\min} < 1$) as a function of the geometric parameters of the rhombic lattice for $n=4$ circular disks.
(d) Maximal achievable HPR as a function of the complex-valued disk permittivity within the sub-diffraction spectral range ($\w/cG_{\min} < 1$). For each permittivity value corresponds to a certain optimal lattice parameters.
(e) $S$-matrix norm, Eq. \eqref{eq:Snorm}, numerically evaluated in the complex-frequency plane near the resonant HPR frequency, revealing a set of poles corresponding to the QNMs of the rhombic metasurface.
}
\label{fig2}
\end{figure*}

We begin our analysis of rhombic metasurfaces with maximization of HP intensity reflection $|r_{RR}|^2$ in an array of dielectric circular disks in air arranged in a rhombic lattice, Fig. \ref{fig2}(a).
The optimization of the geometric parameters of the lattice was conducted for a system of dielectric disks, $n=4$, in air.

\begin{figure*}[t!]
\centering\includegraphics[width=.9\textwidth]{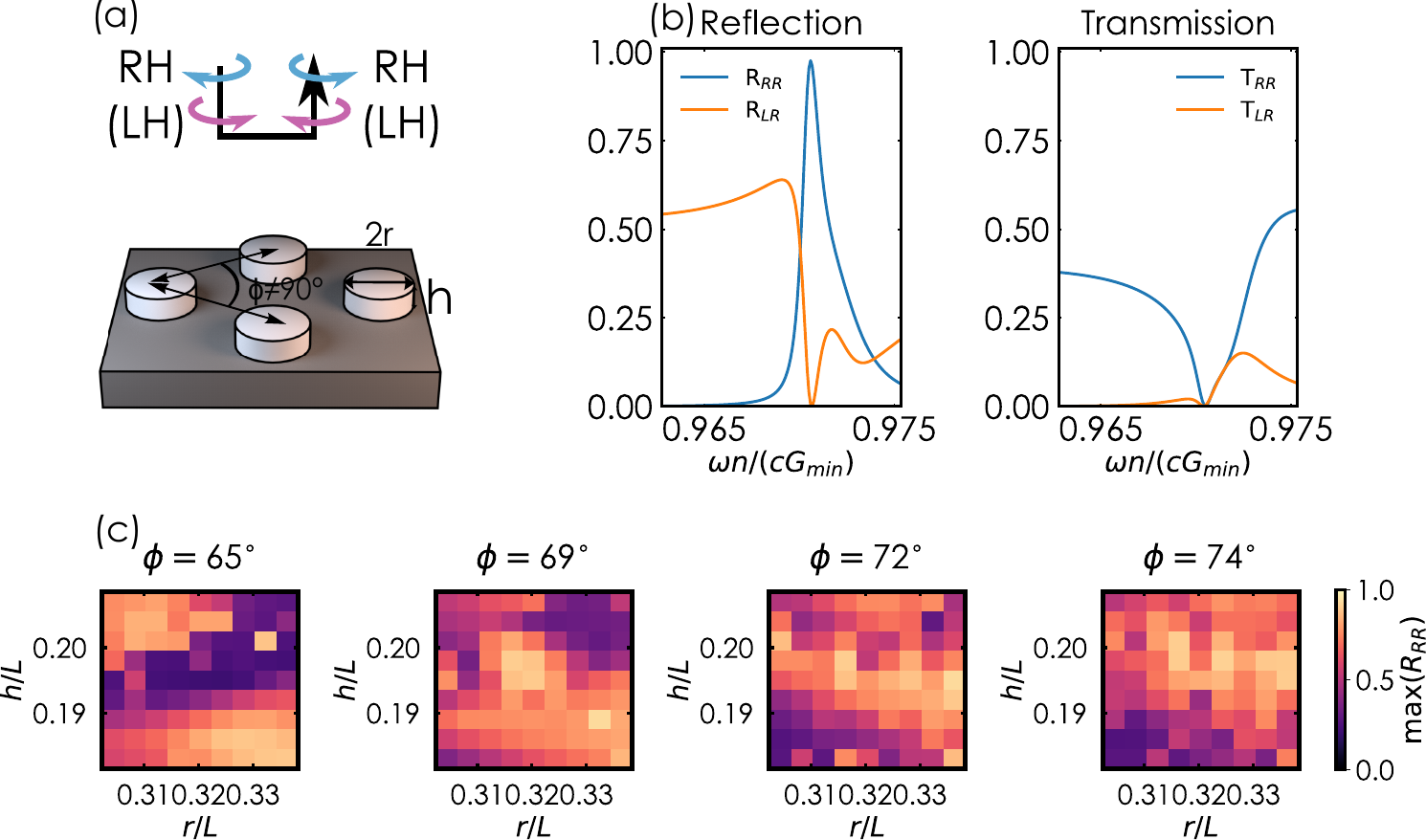}
\caption{
HPR in a rhombic array of dielectric disks on a substrate.
(a) A sketch of the geometry of the system.
(b) Numerically obtained reflection and transmission spectra of an optimal rhombic lattice of dielectric ($n = 4$) disks on a substrate $n_{\mathrm{sub}} = 1.45$ in the circular polarization basis. 
(c) Density plots of maximal achievable HPR ($|r_{RR}|^2$) within the sub-diffraction spectral range ($\w/(cG_{\min}) < 1$) as a function of the geometric parameters of the rhombic lattice for $n = 4$ circular disks on a substrate $n_{\mathrm{sub}} = 1.45$.
}
\label{fig3}
\end{figure*}

Numerical modeling of the structure was performed with Fourier modal method (FMM) using 441 Fourier harmonics \cite{Moharam1995, tikhodeev2002quasiguided, fradkin2019fourier}.
Using the quasi-Newtonian optimization, we achieve near-perfect HPR at normal incidence below diffraction threshold, Fig. \ref{fig2}(b).
Co-polarized reflection and transmission show symmetric Lorentz-like lineshapes near the resonant frequency, while cross-polarized components exhibit more complicated double-peak behavior.

Figure S1 shows the dependence of the HPR magnitude $|r_{RR}|^2$ on the number of harmonics used in FMM simulations at a fixed wavelength corresponding to the resonance for the initial number of harmonics $N_g=441$. 
As the number of harmonics increases, HPR magnitude drops slightly, which is caused by the convergence features of the method (including the displacement of the resonance). 
In a system with a high contrast of dielectric constants of air and the disk material, truncation of the Fourier series leads to slow convergence due to the Gibbs phenomenon.
In the geometry under consideration, the Li factorization rules were not applied, which further slows down convergence. 
Nevertheless, even the initial $N_g = 441$ number of harmonics provides an adequate representation of high HPR.
Figure S2 demonstarates spectra for different number of harmonics. It is observed that for sufficiently large $N_g$
 , the spectra retain their shape; however, their maximum shifts and the amplitude decreases slightly. 

To analyze the robustness of the achieved HP response, we perform sweep of the geometric parameter of the lattice in the vicinity of the optimal point while keeping other parameters fixed, Fig. \ref{fig2}(c). 
For each geometry a new corresponding optimal frequency was found yielding the maximal HPR magnitude $|r_{RR}|^2$ below diffraction threshold for this particular geometry.
The results indicate that the present system is relatively robust against possible geometry variations.

Similar optimization of the geometric parameters was carried out for a series of complex permittivity values containing the imaginary part. 
Figure \ref{fig2}(d) displays these results, where for each permittivity value, we run a separate optimization procedure for the given $\eps$ and determine lattice parameters yielding the highest HPR value at some frequency within the sub-diffraction range.
Permittivity dispersion curves for a number of realistic materials are overlaid with green lines on top. 
As expected, the maximal achievable HPR magnitude drops with growing imaginary part of the dielectric constant. 
However, for the selected materials, many of which exhibit vanishingly small absorption in the infrared range, near-unity HPR $|r_{RR}|^2 \sim 1$ will be achievable.

To uncover the origin of HPR in this structure, we present the $S$-matrix norm expressed as the maximal singular value of the S-matrix:
\begin{equation}
    ||S|| =\sqrt{\max_m \sigma_m},
    \label{eq:Snorm}
\end{equation}
where $\sigma_m$ is the $m$-th singular value of the $S$-matrix.
Figure \ref{fig2}(e) presents the resulting S-matrix norm for the optimal rhombic lattice evaluated numerically in the complex-frequency plane.
One can clearly see a dense collection of poles in the vicinity of the HPR frequency, each corresponding to certain quasi-normal mode (QNMs) of the system \cite{Lalanne2018, TrøstKristensen2021, both2021resonant, sauvan2022normalization}.
These modes likely originate from Mie-like resonances of the individual disks, additionally hybridized by the diffractive lattice interaction \cite{DeAbajo2007, babicheva2019lattice, ustimenko2024resonances, allayarov2025strong}.
This set of QNMs is essentially responsible for the observed HPR behavior.
In principle, the knowledge of these resonant frequencies could allow one to describe the metasurface response analytically on the basis of temporal coupled-mode theory (CMT) \cite{fan2003temporal}; however, the presence of multiple poles seriously complicates such analytical description in this case. 
In the following, we will see how it can be constructed for pure two-mode structures.

\subsubsection{Array of circular disks on a substrate}

Although being a useful toy model for preliminary numerical analysis, a periodic lattice of free-standing disks cannot be implemented experimentally. 
To approach the realistic conditions, we next analyze a similar rhombic array of circular dielectric disks on a dielectric substrate, Fig. \ref{fig3}(a). 
The presence of the substrate violates the horizontal plane of mirror symmetry of the metasurface but preserves the vertical planes, rendering it $C_{2h}$ symmetric.

Figure \ref{fig3}(b) shows numerically obtained optimal reflection and transmission spectra for a rhombic array of circular disks on a glass substrate with $n_\mathrm{sub} = 1.45$
The results again demonstrate near-perfect HPR, $|r_{RR}|^2 \sim 1$, for the optimal set of geometric parameters. 
Analysis of FMM convergence is presented in Fig. S3. 
Similarly to the array of free-standing disks, increasing the number of harmonics slightly reduces the HPR magnitude $|r_{RR}|$ for fixed frequency due a high contrast of dielectric constants of air and the disk material. Nevertheless, using the initial (smaller) number
of harmonics provides an adequate representation of high
HPR.

Similarly to the array of disks in air, the resonant HP response is due to a dense collection of QNMs near the frequency of the HPR.
This is illustrated in Fig. S4 showing the norm of the $S$-matrix of the optimal array of circular disks on the substrate.
Unlike the array of disks in air, however, the geometry sweeps near the optimal lattice parameters uncover highly irregular behavior of the HPR magnitude, Figure \ref{fig3}(c).
For this reason, we turn our attention to the next class of experimentally feasible structures with rhombic lattice.

\subsubsection{Array of circular holes in a membrane}

\begin{figure*}[t!]
\centering\includegraphics[width=.9\textwidth]{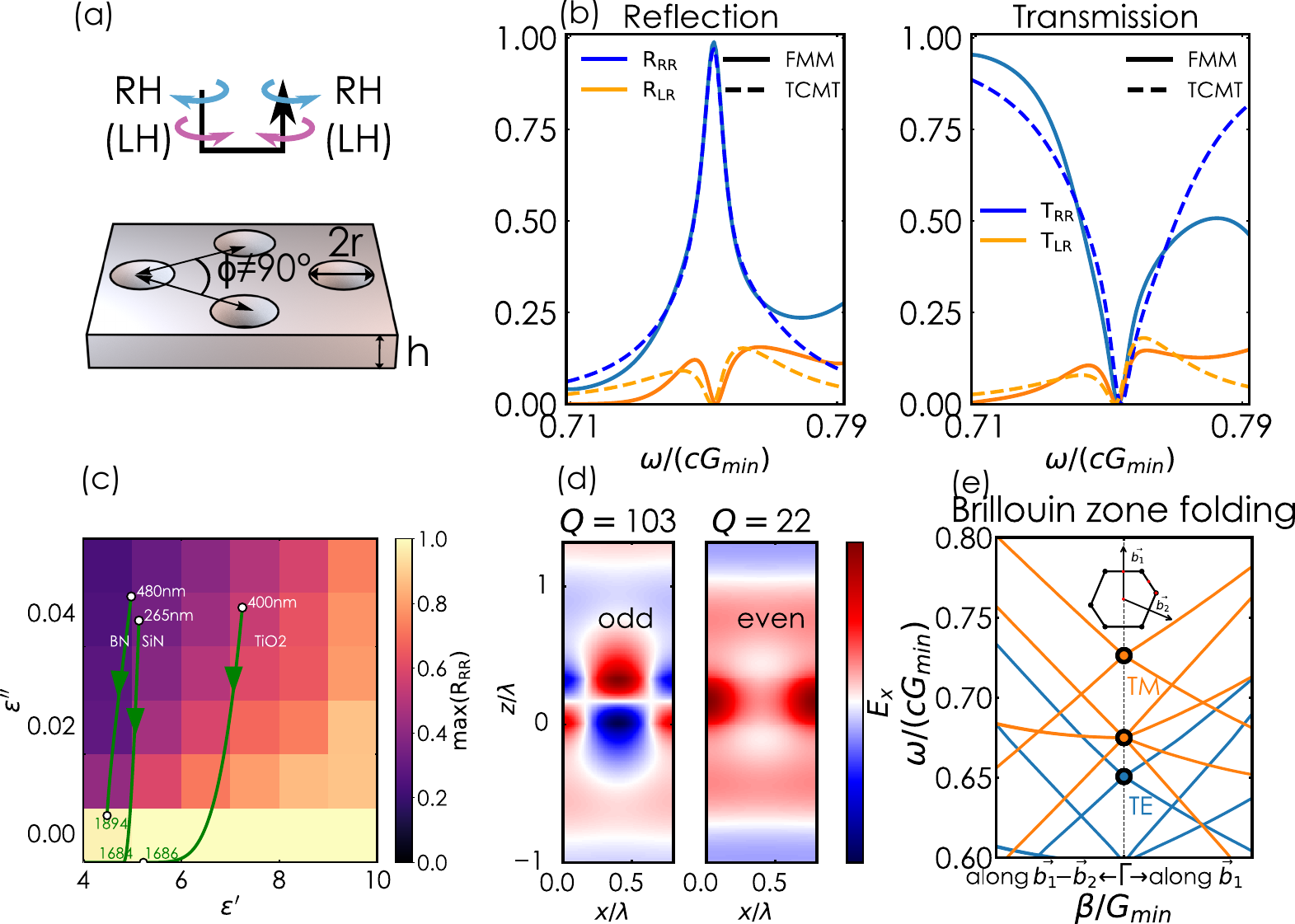}
\caption{HPR in a rhombic lattice of holes within a dielectric plate.
(a) Sketch of the geometry of the system.
(b) Numerically obtained reflection and transmission spectra [presented in the circularly-polarized basis] for an optimal rhombic lattice of holes in a dielectric plate ($n=2$). Solid lines show numerically obtained spectra; dashed lines represent CMT fits.
(c) Map of maximal achievable HPR ($|r_{RR}|^2$) within the sub-diffraction spectral range as a function of the complex-valued permittivity of the plate.
Each pixel corresponds to an optimization of the lattice parameters for a given permittivity.
(d) Electric field distributions for the two dominant quasi-normal modes with opposite parity, identified as the origin of the HPR resonance.
(e) Origin of the resonant modes in the rhombic array of circular holes. Shown is the empty lattice approximation of TE- and TM-polarized guided modes of the unpatterned dielectric film with refractive index $n = 2$, folded into the first Brillouin zone of the rhombic lattice with $G_\mathrm{min}$ being the smallest non-trivial reciprocal lattice vector of the rhombic lattice formed by circular holes. Circles denote the mode frequency at the $\Gamma$-point. Inset illustrates the first Brillouin zone of the metasurface's lattice.
}
\label{fig4}
\end{figure*}

As another example of structures with rhombic symmetry exhibiting HPR we analyze a rhombic lattice of holes within a dielectric membrane, Fig. \ref{fig4}(a).
This system shares the same symmetry properties as the array of free-standing circular disks, ensuring $r_{RR} = r_{LL}$.

Figure \ref{fig4}(b) shows numerically simulated reflection and transmission spectra of a membrane with $n = 2$ with optimal geometric parameters obtained through gradient optimization.
A lower refractive index for this system was chosen since only exceptionally narrow-band resonances were achieved in a series of optimizations with $n=4$. 
The resonant line widths of a membrane with holes are much narrower than those of a dielectric disks array, because the latter represent stronger gratings, their modes are coupled to radiation continuum more effectively, especially when lying on the substrate.
Furthermore, there is an abundance of materials with refractive index approaching $n=2$ in the near-IR and visible range (such as BN, Si3N4, TiO2, to name a few) allowing fabrication of free-standing membranes \cite{Semnani2020}.
The resulting spectra demonstrate near-perfect HPR, $|r_{RR}|^2 \sim 1$, at normal incidence below diffraction threshold.
Figure S5 shows the dependence of the HPR magnitude on the number of harmonics for the membrane system. 

Following a procedure similar to that for the disks array, we investigate the robustness of the HPR regime against material losses by performing a parameter sweep for various complex permittivity values. 
The results, presented in Fig. \ref{fig4}(c), demonstrate that high HPR performance is maintained for materials with low absorption, with the achievable $|r_{RR}|^2$ decreasing as the imaginary part of the permittivity increases.

The $S$-matrix norm, presented in Fig. S7, suggests that the resonant response of the rhombic membrane in the spectral range of interest is governed by a pair of QNMs with opposite parity, whose field profiles are depicted in Fig. \ref{fig4}(d). 
This observation allows us to employ the temporal coupled-mode theory (CMT) in order to analytically described the polarization response of the rhombic metasurface \cite{fan2003temporal, Suh2004}.
In this picture, the response of the metasurface is described by two resonant modes of opposite parity coupled to four (right- and left-handed polarizations incident from two sides of the metasurface) scattering channels.
The general form of CMT equations is:
\begin{equation}
\begin{split}
    \frac{d \mathbf{a}}{dt} &= ( i \mathbf{\Omega}- \mathbf{\Gamma}) \mathbf{a}+ \mathbf{K}^T \sinc, \\
    \sout &= \mathbf{K}\mathbf{a} +\mathbf{C}\mathbf{s}^{+}
\end{split}
\label{Eq_CMT_3}
\end{equation}
where $\sinc =(s^{+}_R, s^{+}_L, s^{+}_R{}', s^{+}_L{}') ^ T$, $\sout =(s^{-}_R, s^{-}_L, s^{-}_R{}', s^{-}_L{}') ^ T$ are the normal input and output waves at two ports, and $\mathbf{a} = (a_1, a_2)$ is the vector of eigenstate amplitudes.
$\mathbf{\Omega}$ is the mode’s resonant frequencies matrix,
$\mathbf{\Gamma}$ is the decay rate operator, and $\mathbf{K}$ is the coupling matrix.
The $\mathbf{C}$ matrix described the non-resonant background scattering processes.

Assuming the mode 1 is even and mode 2 is odd with respect to the horizontal mirror plane, The radiative coupling $4 \times 2$ matrix $\mathbf{K}$ can be written as:
\begin{equation}
    \mathbf{K} = \begin{pmatrix}
        \kappa_{1R} & \kappa_{2R} \\
        \kappa_{1L} & \kappa_{2L} \\
        \kappa_{1L} & - \kappa_{2L} \\
        \kappa_{1R} & - \kappa_{2R} \\
    \end{pmatrix}
\end{equation}
The last two rows are determined by the modal parity with respect to the horizontal mirror plane: the even modes couples symmetrically to the ports, whereas the odd mode changes the sign
(Section S1 of Supporting Information provides detailed derivation of the coupling matrices).
For monochromatic incident field $\sinc(t) = \sinc e^{-i \omega t}$, the scattering matrix takes the form:
\begin{equation}
    \mathbf{S} = \mathbf{C} + \mathbf{K}
    \left[ -( i (\mathbf{\Omega} - \omega \hat{\mathbb{I}})+ \mathbf{\Gamma}) \right]^{-1} 
    \mathbf{K}^T,
\label{Eq_CMT_4}
\end{equation}
which relates the incident and scattered field via 
\begin{equation}
    \sout = \mathbf{S} \sinc.
\end{equation}

\begin{figure*}[t!]
\centering\includegraphics[width=.9\textwidth]{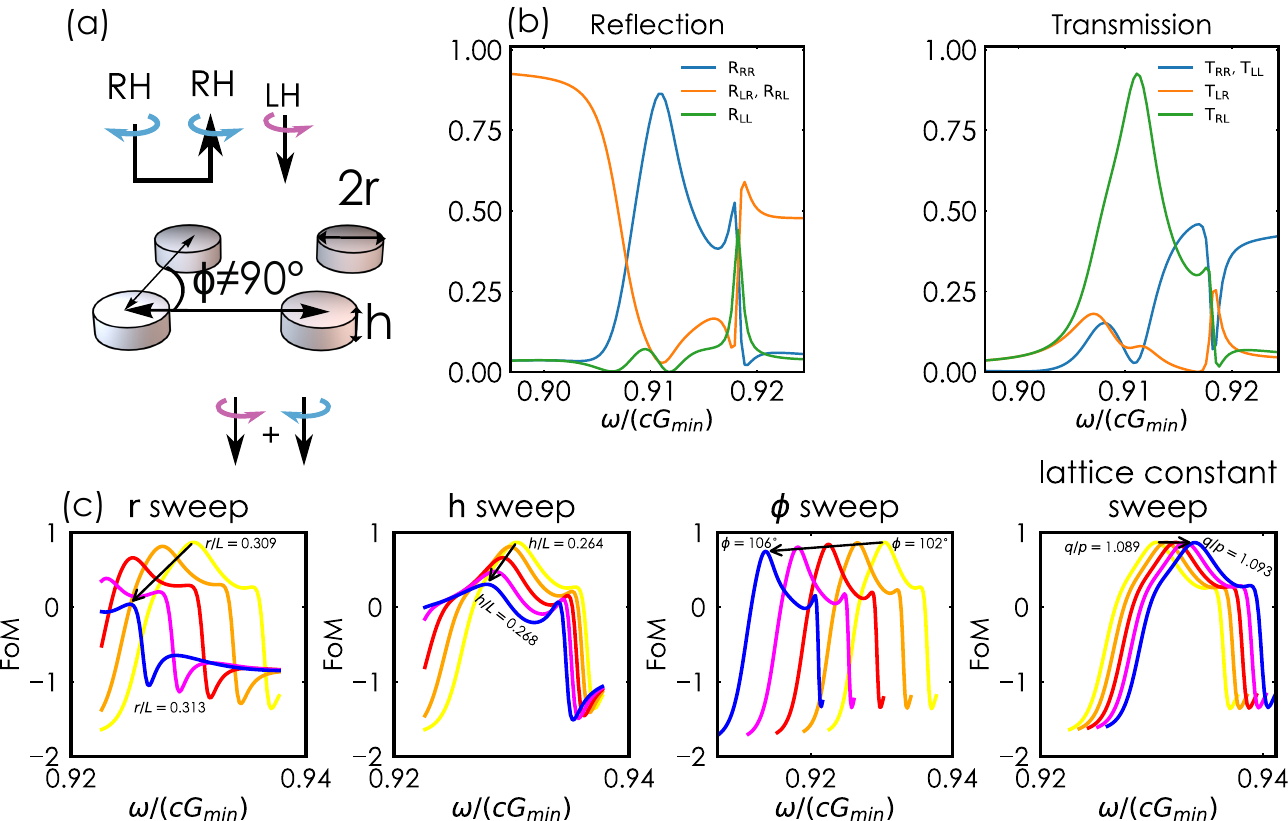}
\caption{Asymmetric HP reflection in a monoclinic lattice of free-standing circular disks.
(a) A sketch of the geometry of the system.
(b) Numerically obtained [circularly-polarized basis] reflection and transmission spectra spectra for an optimized monoclinic lattice of free-standing dielectric disks with $n=4$.
(c) 
The achievable handedness-selective FoM, Eq \eqref{Eq_mono_2}, within the sub-diffraction spectral range of the monoclinic lattice as a function of the lattice geometric parameters.
}
\label{fig5}
\end{figure*}

Figure \ref{fig4}(b) shows fits of numerically simulated power reflection of transmission spectra of the rhombic dielectric membrane with the result of the two-mode CMT model.
The excellent agreement between the numerical results (solid lines) and the CMT fits (dotted lines) validates the applicability of the simple CMT model for the analyzed class of metasurfaces, and allows further prediction of other resonant polarization phenomena in this class of metasurfaces.

The origin of these two even and odd QNMs can be traced back to the underlying guided modes of the unpatterned dielectric film.
The bare film supports a pair of well-known cut-off free TE- and TM-polarized guided modes.
Imposing a periodic pattern on top of the film breaks continuous translational invariance and leads to the emergence of a set of Brillouin zones.
If the periodic perpetuation is sufficiently weak, the presence of this perturbation induces Brillouin zone folding \cite{Overvig2018, Wang2023, Sun2023, dyakov2021photonic}: guided modes below the light line transform into so called guided mode resonances (GMRs) above the light line with nearly identical frequencies:
\begin{equation}
    \beta \to \beta \pm m \mathbf{b}_1 \pm n \mathbf{b}_2.
\end{equation}
Specifically, the pair of TE- and TM-polarized guided modes located at $\beta = G_\mathrm{min}$ appear exactly at the $\Gamma$-point of the newly formed Brillouin zone after Brillouin zone folding [see Fig. \ref{fig4}(e)]. 
These modes are now above the light line and, hence, are available for far-field coupling. In the highly symmetric rhombic lattice, the empty lattice approximation yields two-fold and four-fold degenerate modes in the $\Gamma$-point.


Opposite parity TE- and TM-polarized modes of the homogeneous dielectric film are never degenerate in frequency. However, as one can see from the simple CMT model presented above, the presence of two nearly degenerate modes of opposite parity is necessary to attain near-unity HPR. Thus, one should employ a certain mechanism to shift these two initial guided modes and make them nearly-degenerate. It is exactly the role of the lattice perturbations: their presence not only induces BZF and affects the in-plane momenta of the modes [as shown in Fig.~\ref{fig4}(e)], but also affects the resonant frequencies of the pair of TE and TM modes, rendering them nearly-degenerate.

The precise mechanism works as follows: Perturbation of the initial homogeneous film by introducing the holes lifts the degeneracy of the empty-lattice approximation modes. Thus, spectral positions of the modes of the periodic film are different to those obtained in the empty-lattice approximation. According to the group theory, half of the newly formed modes are symmetry-protected bound states in the continuum while another half are open to the far-field coupling. Among the radiating modes, some of them are even and others are odd. Spectral positions of even and odds modes depend on the perturbation strength set by dielectric permittivity of the holes' material, and these dependencies may intersect. In the vicinity of the intersection point, the grating behaves as HPM, see Fig.~S9 in the Supplemental Materials for details.

Similarly to previous analysis, we also studied the robustness of the HPR regime against perturbations of the geometric parameters of lattice in the vicinity of the optimal point.
The results presented in Fig. S8 suggest that under this variation of the system geometry, a relatively high value of $|r_{RR}|^2$ is retained at the corresponding resonant wavelength. Similar to the disk system, Figure S6 demonstrates spectra for different numbers of harmonics. It is observed that the spectra retain their shape, the amplitude varies less, and the convergence is faster than that of the disks.

\begin{figure*}[t!]
\centering\includegraphics[width=.9\textwidth]{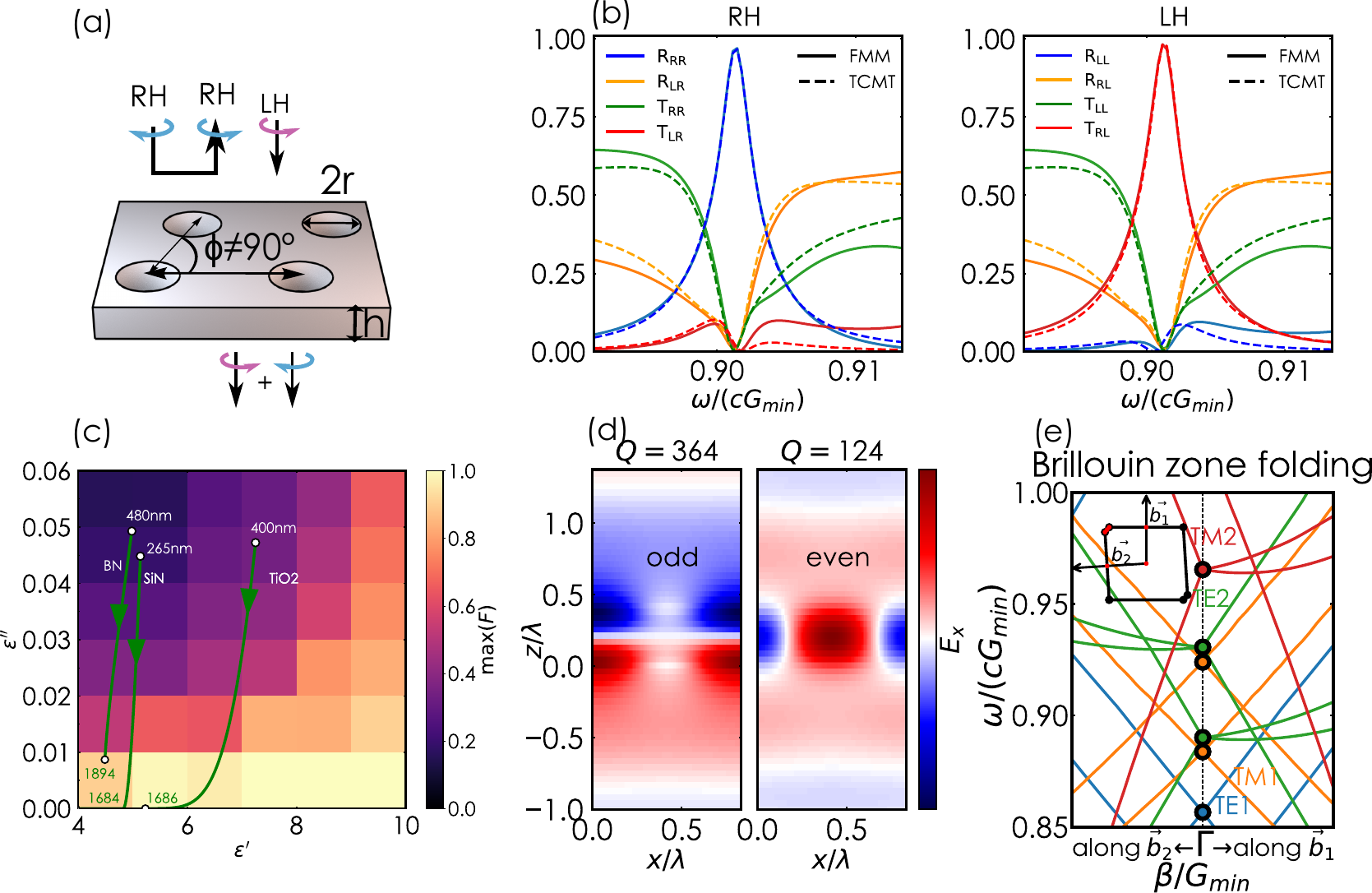}
\caption{Asymmetric HP reflection in a monoclinic lattice of circular holes in a dielectric membrane.
(a) A sketch of the geometry of the system.
(b) Numerically obtained reflection and transmission spectra [circularly-polarized basis] for an optimal monoclinic lattice of holes in a dielectric film ($n=2$). 
Solid lines show numerically obtained spectra; dashed lines represent CMT fits.
(c) Map of the maximal achievable FoM as a function of the complex permittivity of the plate.
(d) Electric field distributions for the two dominant quasi-normal modes with opposite parity, identified as the origin of the studied effects.
(e) Origin of the resonant modes in the monoclinic array of circular holes. Spectrum of the TE- and TM-polarized guided modes of the unpatterned dielectric film with refractive index $n = 2$, folded into the first Brillouin zone of the monoclinic lattice with $G_\mathrm{min}$ being the smallest non-trivial reciprocal lattice vector of the lattice formed by circular holes. Circles denote the mode frequency at the $\Gamma$-point.
Inset illustrates the first Brillouin zone of the metasurface's lattice.}

\label{fig6}
\end{figure*}

\subsection{Monoclinic lattices}

After complete analysis of rhombic metasurfaces, we lower the symmetry of the system by removing the vertical planes of mirror symmetry, and move on to the analysis monoclinic (oblique) lattices having $C_2$ or $C_{2h}$ symmetry.
Thanks to the absence of any vertical plane of mirror symmetry, a sub-diffraction monoclinic metasurface of circular inclusions is capable of completely reflecting light of one handedness and transmitting the other:
\begin{equation}
     r_{RR} = 1, \quad r_{LL} = 0.
\label{Eq_mono_1}
\end{equation}
We aim to to obtain a structure with near-optimal $|r_{RR}|$, and at the same time suppress all other unwanted reflection processes.
To quantify the polarization-selective performance of a monoclinic metasurface we thus introduce the following handedness-selective figure of merit:
\begin{equation}
     \mathrm{FoM} = |r_{RR}|^2 - |r_{RL}|^2 - |r_{LR}|^2 - |r_{LL}|^2.
\label{Eq_mono_2}
\end{equation}
which takes the values in the range range from -2 (unwanted performance) to 1 (the optimal performance).

Two-dimensional monoclinic lattice is defined by the real-space translation vectors
\begin{equation*} \mathbf{a}_1 = a(1,0), \qquad
\mathbf{a}_2 = b(\cos\phi,\;\sin\phi), \end{equation*}
where $a$ and $b$ are the lengths of the basis vectors, $  a \neq b  $, and $\phi$ is the angle between them. The corresponding reciprocal lattice vectors are given by
\begin{equation*}  
\mathbf{b}_1 = \frac{2\pi}{a}\left(1,\;-\cot\phi\right), \qquad
    \mathbf{b}_2 = \frac{2\pi}{b\sin\phi}\,(0,\;1),
\label{Eq_mono_3}
\end{equation*}
Magnitudes of the few relevant lowest-order reciprocal vectors are given by:
\begin{equation}\begin{split}
    |\mathbf{G}_{\pm1,0}| &= \frac{2\pi}{a\sin\phi}, \\
    |\mathbf{G}_{0,\pm1}| &= \frac{2\pi}{b\sin\phi}, \\
    |\mathbf{G}_{1,1}| &= |\mathbf{G}_{-1,-1}| = \frac{2\pi}{\sin\phi}\sqrt{\frac{1}{a^2} + \frac{1}{b^2} - \frac{2\cos\phi}{ab}}, \\
    |\mathbf{G}_{1,-1}| &= |\mathbf{G}_{-1,1}| = \frac{2\pi}{\sin\phi}\sqrt{\frac{1}{a^2} + \frac{1}{b^2} + \frac{2\cos\phi}{ab}}.
\end{split}
\end{equation}
Similar to the rhombic lattice, at normal incidence the sub-diffraction regime corresponds to frequencies below the first diffraction threshold:
\begin{equation*}  
    \w /c < G_\mathrm{min}
\end{equation*}
where
\begin{equation*}  
    G_\mathrm{min} \equiv \min_{(m,n)\neq(0,0)} |\mathbf{G}_{m,n}|
\label{Eq_mono_5}
\end{equation*}
denotes the smallest non-trivial reciprocal lattice vector
of the monoclinic lattice

\subsubsection{Array of circular disks in air}

Similarly to the analysis of rhombic lattices, we begin with the analysis of a monoclinic array of circular dielectric disks, $n = 4$, in air, Fig. \ref{fig5}(a).
Optimization of the FoM of interest, \eqref{Eq_mono_2}, was performed with respect to the following geometric parameters: the angle of the lattice, the ratio of the base lengths of the vectors, the radii and heights of objects.
Figure \ref{fig5}(b) shows exemplary reflection and transmission spectra of a near-optimal system.
The spectra demonstrate near-optimal HP reflection with $|r_{RR}|^2 \approx 0.96$ and $\mathrm{FoM} \approx 0.93$ below the diffraction threshold. However, the spectral response exhibits a complex behavior, featuring multiple peaks and sharp dips.
These complex resonant features again can be attributed to the presence of several interfering resonances in the relevant frequency range, see Fig. S10.

Figure \ref{fig5}(c) shows a series of geometry sweeps for handedness-selective $\mathrm{FoM}$, Eq. \eqref{Eq_mono_2}, with respect to the circular disk radius, height, and the lattice constant. 
We opted for separate one-dimensional plots because they provide a clearer visualization of the individual impact of each geometric parameter on the FoM, avoiding the complexity of 2D false-color maps. 
This approach allows for a direct comparison of how each parameter tunes the spectra. 
Overall, FoM values around 0.93 can be obtained below diffraction at reasonable lattice parameters, although values exceeding 0.95 are not achieved within the scanned ranges of radius and height. 
One can see that the FoM progressively varies non-monotonically upon change of the disk radius, while the response to height variations is relatively weaker. 

Overall, the system is relatively stable against geometry variations.
However, the spectra of such a system have rather complex behavior for the same reason as with rhombic arrays of disks, as they are the result of overlap between numerous Mie-like modes of the single disk; additionally, such a system is hardly implemented experimentally.
Thus, analogous to Section II.A.3, we move on to the analysis of a monoclinic array of circular holes in a dielectric membrane.

\subsubsection{Array of circular holes in a membrane}

\begin{figure*}[t!]
\centering\includegraphics[width=1.0\textwidth]{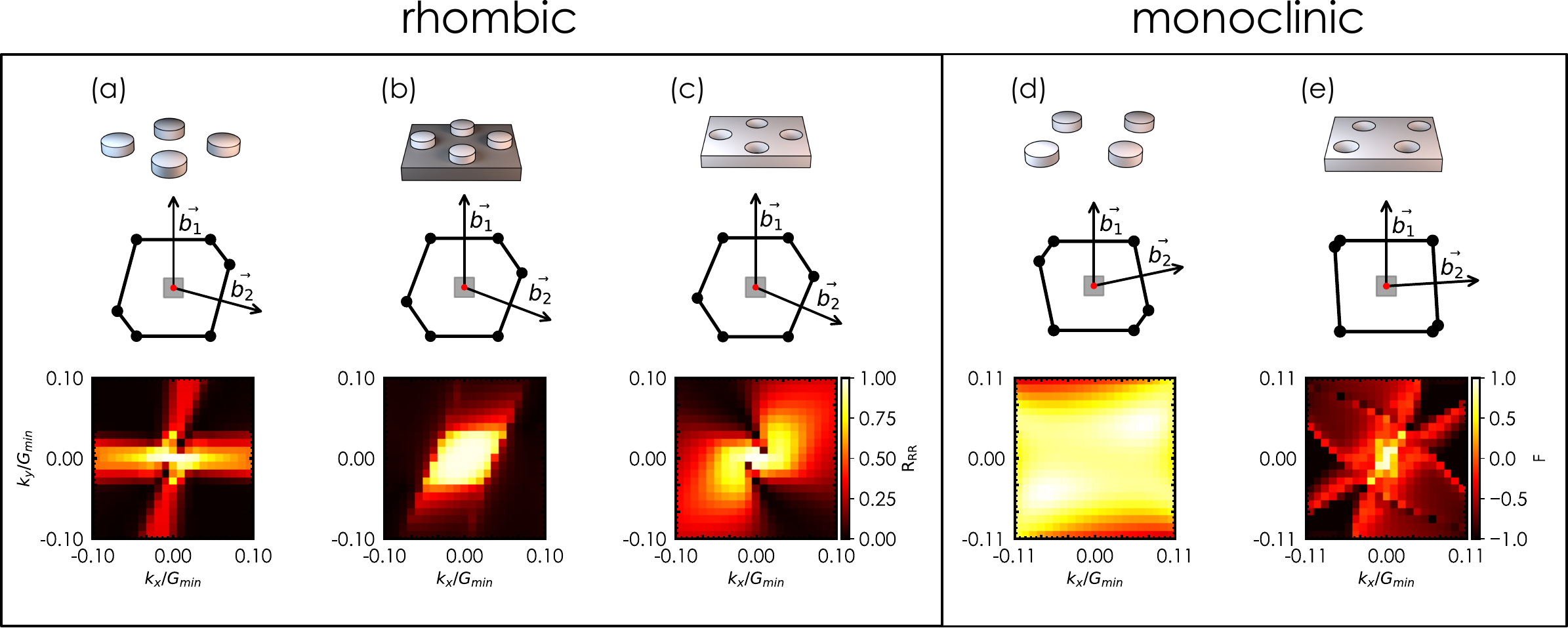}
\caption{Analysis of HPR at oblique incidence on rhombic and monoclinic lattices.
Top: Schematics of the first Brillouin zones for (a) - (c) rhombic and (d), (e) monoclinic systems with corresponding reciprocal lattice vectors $\mathbf{b}_1$ and $\mathbf{b}_2$.
Bottom: corresponding density plots of (a-c) HPR magnitude $|r_{RR}|^2$ for rhombic lattices and (d, e) asymmetric $\mathrm{FoM}$ for monoclinic lattices as a function of in-plane wave vector at a fixed frequency corresponding to the optimal HP response at normal incidence.}
\label{fig7}
\end{figure*}

Similarly to the monoclinic array of circular disks, we optimize the geometry of the monoclinic lattice of circular holes in a $n = 2$ dielectric membrane to maximize the handedness-selective FoM, Fig. \ref{fig6}(a). 
Figure \ref{fig6}(b) shows exemplary reflection and transmission spectra of a near-optimal system.
The spectra clearly demonstrate nearly optimal ($|r_{RR}|^2 \approx 1$, $|r_{LL}| \approx 0$) asymmetric HPR behavior with a simple Lorentz-like resonant reflection line, suggesting this operation can be attributed to a pair of nearly-degenerate opposite parity modes, just like in the case of a rhombic membrane.

Next we analyze the robustness of the asymmetric HPR regime against material losses by performing a sweep of complex permittivity of the membrane and evaluating the maximum achievable handedness-selective FoM in the sub-diffraction regime, Fig. \ref{fig6}(c).
Similarly to the analogous results for the rhombic membrane, near-optimal asymmetric HPR performance is maintained for materials with low absorption, achievable for BN, Si3N4, TiO2 in the near-IR range.

Similarly to the rhombic membrane, the resonant HP response of the monoclinic membrane originates from the presence of two nearly-degenerate modes of opposite parity (see $S$-matrix norm presented in Fig S11), whose electric field profiles are shown in Fig. \ref{fig6}(d).
The spectral response thus can be well described by the same two-mode CMT model.
The $S$-matrix will be given by the same Eq. \eqref{Eq_CMT_4}. 
Since the structure still preserves the horizontal mirror plane, the coupling matrix retains the same general form as in the rhombic case:
\begin{equation}
    \mathbf{K} = \begin{pmatrix}
        \kappa_{1R} & \kappa_{2R} \\
        \kappa_{1L} & \kappa_{2L} \\
        \kappa_{1L} & - \kappa_{2L} \\
        \kappa_{1R} & - \kappa_{2R} \\
    \end{pmatrix}
\end{equation}
In contrast to the rhombic lattice, however, the absence of the vertical mirror plane does not impose the additional symmetry relations between the couplings to opposite circular polarizations. Therefore, the monoclinic response is described by the same two-mode CMT, but with a different set of fitting parameters; further details are given in Section~S1 of the Supporting Information.
The two-mode CMT model reproduces very well the numerical reflection and transmission spectra, Fig. \ref{fig6}(b). 

The origin of the two nearly-degenerate resonances of a monoclinic membrane is analogous to those of a rhombic membrane: these are GMRs originating from the Brillouin zone folding of the true guided modes of the dielectric membrane [Fig.~\ref{fig6}(e)]. 
Introducing the periodic holes perturbs the homogeneous film, lifting the degeneracy of the empty-lattice modes. Some of the resulting modes become symmetry-protected bound states in the continuum, while the other couple to the far field, splitting into even and odd radiating modes. When their spectral positions cross as a function of the hole’s dielectric permittivity, the system behaves as an HPM.


\subsection{Oblique incidence}

To complete our study, we analyze the performance of all five classes of low-symmetry (rhombic and monoclinic) metasurfaces at oblique incidence.
We examine the dependence of the target functions ($|r_{RR}|^2$ for the rhombic and $\mathrm{FoM}$, Eq. \eqref{Eq_mono_2}, for the monoclinic structures) at the normal-incidence resonant frequency as a function of the in-plane wave vector $\mathbf{k}_\parallel = (k_x, k_y)$. 

Figure \ref{fig7} summarizes the results by presenting numerically calculated target functions ($|r_{RR}|^2$ for rhombic and $\mathrm{FoM}$ for monoclinic lattices) for previously obtained optimal geometries of (a) rhombic arrays of free-standing disks, (b) rhombic arrays of disk on a substrate, (c) rhombic arrays of circular holes in a membrane, (d) monoclinic arrays of free-standing disks, and (e) monoclinic arrays of circular holes in a membrane.
The results are presented as a function of the in-plane wave vector at a fixed frequency (at which the system demonstrates the optimal HP response at normal incidence).
Each panel also visualizes the corresponding Brillouin zone, with the analyzed region of in-plane wave vectors indicated by the gray square. 

Overall, pronounced dependence of HP reflection on the incidence direction is evident.
The handedness-selective $\mathrm{FoM}$ for the optimal monoclinic array of holes in the dielectric membrane shows particularly strong dependence, Fig. \ref{fig7}(e). 
This behavior can be attributed to the large $Q$-factors ($> 300$) of the modes involved in the response of the monoclinic membrane [see Fig. \ref{fig6}(d)].
Another reason of this strong angular dependence could be the pronounced dispersion of the folded guided modes of the dielectric film.
At the same time, the handedness-selective $\mathrm{FoM}$ in the monoclinic array of circular disks shows the least pronounced angle dependence among the five exemplary analyzed metasurfaces, Fig. \ref{fig7}(d).

The observed strong angle dependence raises the question of developing handedness-preserving reflective metasurfaces with broadband and omnidirectional response, possibly by lowering the $Q$-factors of the resonant modes involved in the HP response.
Recent theoretical \cite{Dyakov2025} and experimental \cite{Salakhova2026} demonstrations show a viable route towards desired broadband and wide-angle HP response, yet only in a mirror-symmetric case, $r_{RR} = r_{LL}$. The question of designing a broadband and wide-angle asymmetric HP metasurface thus remains open.

\section{Conclusions}

To conclude, we have presented detailed numerical and theoretical analysis of handedness-preserving reflection via a family of low-symmetry rhombic and monoclinic periodic lattices composed of high-symmetry non-chiral meta-atoms.
Full-wave numerical simulations based on the Fourier modal method yield a series of optimal geometries exhibiting near-perfect HP reflection around normal incidence.
The analysis shows that the rhombic lattices of holes in a dielectric film proves to be the most robust against geometric perturbations.
The obtained handedness-preserving resonant response of rhombic and monoclinic arrays of circular holes in a dielectric film is well described by the analytical temporal coupled-mode theory.
All studied systems -- except for the free standing monoclinic arrays of circular disks -- exhibit rather strong dependence of handedness-preserving reflection on the incidence direction. 
Recent experimental results on C$_{2v}$ symmetric HP metasurfaces on Bragg mirrors suggest a possible strategy to improve the wide-band and wide-angle response \cite{Salakhova2026}.
Our results highlight the potential of low-symmetry metasurfaces of simple circular inclusions for handedness-preserving reflection, where conventional metasurface designs face fundamental limitations.


\section{Acknowledgements}

We thank Maxim Gorkunov, Yuri Kivshar, and Ivan Toftul for stimulating discussions.
The work was supported by the Ministry of Science and Higher Education of the Russian Federation (FSMG-2024-0014).
The authors acknowledge support from Russian Science Foundation (Grant No. 25-12-00454).

\bibliography{Lattices.bib}

\end{document}